\documentclass[aps,prd,preprint,superscriptaddress,nofootinbib]{revtex4-2}
\usepackage{amsmath,amssymb,color,graphics}
\usepackage{bm}
\usepackage{hyperref} 

\definecolor{dark-green}{rgb}{0,0.7,0}
\definecolor{dark-blue}{rgb}{0,0.2,0.5}
\definecolor{med-blue}{rgb}{0,0.7,1}
\definecolor{mblue}{rgb}{0,0.2,1}
\definecolor{cnc}{rgb}{0.8,0,0}
\definecolor{light-red}{rgb}{1,0.8,0.8}
\definecolor{dark-yellow}{rgb}{1,0.8,0}
\definecolor{light-blue}{rgb}{0.8,0.9,1}
\definecolor{verylight-blue}{rgb}{0.93,0.95,1}
\definecolor{light-yellow}{rgb}{1,0.9,0.8}
\definecolor{grey}{gray}{0.88}

\begin{document}
\title{Nonmetricity and hypermomentum: on the possible violation of Lorentz invariance}
\author{Yuri N. \surname{Obukhov}}
\email{obukhov@ibrae.ac.ru} 
\affiliation{Theoretical Physics Laboratory, Nuclear Safety Institute,
  Russian Academy of Sciences, B.Tulskaya 52, 115191 Moscow, Russia}

\author{Friedrich W. \surname{Hehl}}
\email{hehl@thp.uni-koeln.de}\homepage{http://www.thp.uni-koeln.de/gravitation/mitarbeiter/hehl.html}
\affiliation{Faculty of Mathematics and Natural Sciences, Institute for Theoretical Physics, University of Cologne, 50923 Cologne, Germany}

\begin{abstract}
We outline how the symmetry groups of spacetime are interpreted in a gauge-theoretic approach. Specifically, we focus on the hypermomentum concept and discuss the hyperfluid, that appropriately generalizes the perfect (Euler) fluid of general relativity to the case of continuous media with microstructure. We demonstrate that a possible violation of Lorentz invariance is most adequately understood by means of non-vanishing nonmetricity of a metric-affine geometry of spacetime.
\end{abstract}
\maketitle
\begin{footnotesize}
\tableofcontents
\end{footnotesize}  

\section{Introduction}

Modern understanding of gravitational phenomena is based on the concepts of spacetime geometry. In a broad sense, the theory of gravity deals with the dynamics of the geometry that arises from the motion of matter. In his paper ``Geometry and experience'' \cite{Einstein:2002}, Einstein advocated to view the geometry as a natural science (actually he had regarded it as the most ancient branch of physics) and summarized this approach in the statement which can hardly be disputed\footnote{The original German text \cite{Einstein:1921} reads: ``Die Frage, ob dieses Kontinuum euklidisch oder gem\"a\ss\ dem allgemeinen Riemannschen Schema oder noch anders strukturiert sei, ist nach der hier vetretenen Auffassung eine eigentlich physikalische Frage, die durch die Erfahrung beantwortet werden mu\ss, kein Frage blo{\ss}er nach Zweckm\"a{\ss}igkeitsgr\"unden zu w\"ahlender Konvention.''}: {\it ``According to the view advocated here, the question whether this continuum has a Euclidean, Riemannian, or any other structure is a question of physics proper which must be answered by experience, and not a question of a convention to be chosen on grounds of mere expediency.''}

As is well known, the non-gravitational sector of physics (that encompasses the electromagnetic, weak, and strong interactions) is essentially explained in the framework of a Yang-Mills type gauge-theoretic approach \cite{ORaifeartaigh:1978,Mack:1979,Chaichian:1984}. The latter is based on the fundamental symmetry groups acting in internal spaces. Seeking for an extension of the gauge principle to the gravitational case, one naturally comes from internal symmetries to the `external' spacetime symmetry groups. Thereby, in the gauge-theoretic paradigm, Einstein's question on an experimental probing of the geometrical structure of spacetime is converted into a question of an experimental determination of the underlying fundamental symmetry of the spacetime.

With this ambitious goal in mind, we have then to think about the appropriate physical tools which are suitable to probe the spacetime geometry. For example, neutral test particles do not feel the presence of electric and magnetic fields, one needs test particles that carry electrical charge. Similarly, exploration of the spacetime structure requires physical matter with adequate ``gravitational charges''. Remarkably, the nature of such matter is perfectly determined from the gauge-theoretic principles in terms of the Noether currents corresponding to the underlying spacetime symmetry.

It is our pleasure to prepare a paper for the volume devoted to the memory of {\it Ruben Aldrovandi}, who made important contributions to the development of geometrical methods in physics. He consistently shared Einstein's view, aptly noticing that {\it ``the marriage between geometry and physics is best perceived when we notice that the geometry of Nature is probed by the physical dynamics''} \cite{Aldrovandi:2017}. Both of us were influenced by this remarkable book.

\section{General relativity}

The arena of {\it classical mechanics} is the Galilei-Newton spacetime with the Galilei group as group of motion. Maxwellian {\it electrodynamics} could not be accommodated in such a spacetime. The group of motion had to be generalized to the Lorentz-Poincar\'e group residing in the Minkowski spacetime of special relativity. As soon as {\it gravitation} is taken into account additionally, Einstein demonstrated that the Minkowski geometry of special relativity had to be generalized to the Riemannian geometry of general relativity (GR). This whole development has been conclusively reviewed in Einstein's Princeton lectures of 1921 under the title of ``The Meaning of Relativity'' \cite{EinsteinMeaning}.
 
In the context of this development of classical mechanics to electrodynamics and, eventually, to gravity, the Galilei-Newton geometry of spacetime was first generalized to Minkowski geometry of special and then to the Riemannian geometry of GR. Apparently, if spacetime and its underlying group of motion is adapted to more advanced experimental results, the {\it geometry} of spacetime has to be appropriately generalized. Thus, naively following the course of history, if a spacetime symmetry is violated, Lorentz invariance, for instance, then a more general type of spacetime geometry is expected to unfold.

When Einstein developed general from special relativity, he described matter phenomenologically  as a  {\it  perfect fluid,} see \cite[Eq.(51)]{EinsteinMeaning}. The dynamics of such a continuous medium is characterized by an average 4-velocity vector field $u^i$ (normalized as $g_{ij}u^iu^j = c^2$) and a scalar energy density field $\rho$. If necessary, this can be straightforwardly generalized by including elastic stresses. It is a distinguishing feature of GR that there is an intimate relation between the geometry of spacetime and matter. If we denote the matter Lagrangian density of the perfect fluid by $\mathfrak{L}_{{\rm mat}} = \sqrt{-g}\,L_{\rm mat}$, then a variation of the Riemann metric $g_{ij}$ (with $i,j=0,1,2,3$) yields the covariantly conserved and symmetric energy-momentum tensor $\sigma^{ij}$ of the perfect fluid, see Hilbert \cite{Hilbert}:
\begin{equation}\label{sigH}
\sigma^{ij}=\sigma^{ji} = -\,{\frac 2{\sqrt{-g}}}{\frac {\delta\mathfrak{L}_{{\rm mat}}}{\delta g_{ij}}},
\qquad \stackrel{\{\}}{\nabla}_j \sigma^{ij}=0.
\end{equation}
Here $g = \det g_{ij}$, and $\stackrel{\{\}}{\nabla}$ denotes the covariant derivative with respect to the Riemannian (Levi-Civita) connection
\begin{equation}\label{LC} 
\{{}_k{}^i{}_j\} = {\frac 12}g^{il}\left(\partial_jg_{kl} + \partial_kg_{jl} - \partial_lg_{kj}\right).
\end{equation}
The Noether (canonical) energy-momentum current of the {\it pressure-less} fluid (= dust) has a particularly simple form
\begin{equation}\label{SigF}
\Sigma_k{}^i = u^i{\mathcal P}_k,\qquad {\mathcal P}_k = {\frac {\rho}{c^2}}\,u_k, 
\end{equation}
and energy-momentum conservation law (\ref{sigH}) is supplemented by the angular momentum conservation
\begin{equation}\label{angF}
\Sigma_{[ij]} = 0,\qquad \Sigma_{(ij)} = \sigma_{ij}.
\end{equation}
In the words of Weyl \cite[p.~237]{Weyl1952}: {\it ``The general theory of relativity alone, which allows the process of variation to be applied to the metrical structure of the world, leads to the true definition of energy''}. Accordingly, the properties of matter are directly reflected in the geometry of spacetime---and vice versa. This already gives us an inkling that a possible refined phenomenological description of matter could have implications for the geometry of spacetime in question.

\section{Metric-affine geometry of spacetime}

Already the Galilei-Newton spacetime of classical mechanics carries a metric-affine geo\-metry with a 4-dimensional degenerate metric and a linear connection, see Weyl \cite{Weyl1952}, Havas \cite{Havas}, and, particularly, Kopczy\'nski and Trautman \cite[Chap.~3]{Kopczynski}. This dichotomy of metric $g$ and linear connection $\Gamma$ runs through the whole of relativistic physics.  In general relativity theory, however, the connection is subordinate to the spacetime metric: The Levi-Civita (Christoffel) connection (\ref{LC}) can be expressed in terms of the metric and its first derivatives alone, $\Gamma_{kj}{}^i = \{{}_k{}^i{}_j\}$.

Still, it is worthwhile to mention that Einstein clearly understood the different physical statuses of the metric and connection. Because of the principle of inertia,  {\it ``...the essential achievement of general relativity, namely to overcome ‘rigid’ space (ie the inertial frame), is {\it only indirectly} connected with the introduction of a Riemannian metric. The directly relevant conceptual element is the ‘displacement field’ ($\Gamma^l_{ik}$), which expresses the infinitesimal displacement of vectors. It is this which replaces the parallelism of spatially arbitrarily separated vectors fixed by the inertial frame (ie the equality of corresponding components) by an infinitesimal operation. This makes it possible to construct tensors by differentiation and hence to dispense with the introduction of ‘rigid’ space (the inertial frame). In the face of this, it seems to be of secondary importance in some sense that some particular $\Gamma$ field can be deduced from a Riemannian metric...'',} as Einstein formulated\footnote{Translation from the German original \cite{Einstein1955} by F. Gronwald, D. Hartley, and F. W. Hehl.} in one of his last publications \cite{Einstein1955}, see also Einstein \cite[Appendix II]{EinsteinMeaning} and Schr\"odinger \cite{Schrodinger1952}.

Thus, the straightforward generalization of the Riemannian spacetime is the metric-affine spacetime with a (symmetric\footnote{Einstein as well as Schr\"odinger used in their unified field theories also an {\bf}asymmetric metric. Since the antisymmetric part of the metric has no generic geometrical interpretation, we restrict ourselves to its symmetric part. Moreover, they used the metric-affine framework in the context of a unified theory for gravity and electrodynamics (and possibly meson fields \cite{Schrodinger1952}). We, however, opt for a dualistic theory like GR in which matter interacts with the geometry of spacetime.}) metric $g_{ij}$ and an asymmetric linear connection $\Gamma_{ij}{}^k$; we use here Schouten's \cite{Schouten1954} conventions. In general, these geometrical objects are completely independent, and the spacetime geometry is exhaustively characterized by the tensors of {\it curvature}, {\it torsion}, and {\it nonmetricity}, respectively:
\begin{align}
R_{kli}{}^j &:= \partial_k\Gamma_{li}{}^j - \partial_l\Gamma_{ki}{}^j + \Gamma_{kn}{}^j \Gamma_{li}{}^n - \Gamma_{ln}{}^j\Gamma_{ki}{}^n,\label{curv}\\
T_{kl}{}^i &:= \Gamma_{kl}{}^i - \Gamma_{lk}{}^i,\label{tors}\\ \label{nonmet}
Q_{kij} &:= -\stackrel{\Gamma}{\nabla}_k\!g_{ij} = - \,\partial_kg_{ij} + \Gamma_{ki}{}^lg_{lj} + \Gamma_{kj}{}^lg_{il}.
\end{align}
Here $\stackrel{\Gamma}{\nabla}$ denotes the covariant derivative with respect to the connection $\Gamma$.

A metric-affine space is called a {\it Riemann-Cartan space} (or a space with a metric-compatible connection), if its nonmetricity is vanishing: $Q_{kij}=0$. It is called a Riemann space, provided its torsion vanishes additionally: $T_{ij}{}^k=0$. Then the linear connection reduces to (\ref{LC}). 

To paraphrase Weyl's statement on the metric and the energy-momentum of matter, we may say that only a metric-affine spacetime which allows the process of the variation to be applied to the affine structure of the world, $\Gamma_{ij}{}^k\rightarrow \Gamma_{ij}{}^k+\delta \Gamma_{ij}{}^k$, leads to the true definition of the {\it hypermomentum} of matter:
\begin{equation}\label{D}
\Delta^i{}_j{}^k = {\frac {\delta{L}_{{\rm mat}}}{\delta \Gamma_{ki}{}^j}}.
\end{equation}
The meaning of this current $\Delta^i{}_j{}^k$ will be discussed below. A fluid with non-vanishing hypermomentum we will call a hyperfluid. We follow here the investigations of Obukhov and Tresguerres \cite{Obukhov:1993}, see also \cite{Obukhov:2023}.

\section{Gauge theories of gravitation as a unifying framework}

The interdependence between the fundamental geometrical objects of spacetime, the gravitational field {\it potentials} (metric and connection) and the physical {\it sources}\footnote{See also Schwinger's \cite{Schwinger2} source concept.} of gravity (energy-momentum and hypermomentum currents of matter), can be readily understood in terms of a gauge-theoretic framework for gravity.

\'Elie Cartan's idea in the early 1920s of {\it gauging} the (4+6) parameter {\it Poincar\'e} (inhomogeneous Lorentz) group with mass $m$ and spin $s$ as the corresponding currents came eventually to fruition by the investigations of Sciama and Kibble \cite{Sciama:1962,Kibble:1961} during the early 1960s. The simplest gauge theory of gravity is the Einstein-Cartan(-Sciama-Kibble) theory of gravity, an experimentally viable generalization of Einstein's theory. Besides the energy-momentum of matter, also its spin angular momentum acts as source of gravity. The perfect fluid in Einstein's theory is substituted by a spin fluid, see Halbwachs \cite{Halbwachs,Halbwachs2}. 

The Einstein-Cartan theory \cite{Trautman:2006,Primer} leads only to minute deviations from general relativity at extremely high matter densities ($\approx 10^{57}$kg/m$^3$), which may only be relevant in the early cosmos. The corresponding historical and epistemological development is most convincingly described by O'Raifeartaigh \cite[Chap.~3]{O'Raifeartaigh} and by Cao \cite[Secs.~11.2, 11.3]{Cao}. The relevant important original papers are reprinted and explained in Blagojevi\'c et al.\cite{Blagojevic:2013xpa}.

Encouraged by the successful gauging of the Poincar\'e group,  this concept was extended to the 4-dimensional general affine group $A(4,R)=T(4)\rtimes GL(4,R)$, see \cite{Hehl:1976kx,Hehl:1978cb,Hehl:1994ue}, and \cite[Chap.~9]{Blagojevic:2013xpa}. Thereby local Lorentz invariance is violated by the emergence of the nonmetricity $Q_{kij}$, a geometrical object of spacetime. Of course, it is a question to experimental physics whether in some era of the cosmos really a violation of Lorentz invariance emerged. So far no Lorentz violation has been experimentally found. But numerous frameworks have been constructed for the accommodation of such a possible effect.

\section{Possible violation of Poincar\'e-Lorentz invariance}

The Poincar\'e group $T(4)\rtimes SO(1,3)$ and the affine group $T(4)\rtimes GL(4,R)$ both embrace the translation group $T(4)$. The latter relates in each case {\it two} neighboring points of spacetime to each other. Should the translational symmetry be violated, then we expect a fundamental change in the underlying spacetime geometry.  In fact, the ``square roots'' of the 4 translation generators are believed to be complemented by the 4 additional supersymmetry generators of simple supergravity, see \cite[Chap.~12]{Blagojevic:2013xpa}. Thus, there are well-developed methods in supergravity for treating this case. In this article, we will assume  that translation invariance is left untouched and we only discuss possible violations of (homogeneous) Lorentz invariance.

The Lorentz group $SO(1,3)$---in contrast to the translation group $T(4)$---acts at one and only one point in spacetime. Accordingly, it is possible to treat Lorentz invariance like an internal symmetry group $U(1), SU(2),SU(3)$ etc. Below, we shall shortly come back to this possibility. If we understand, however, the Lorentz group as an external group---and this corresponds to our present state of knowledge---then the violation of the Lorentz symmetry means that the group $SO(1,3)$ expands to the general linear group $GL(4,R)$:  $SO(1,3)\rightarrow GL(4,R)$. Physically, the Lorentz symmetry violation is manifest in that the light cone looses a status of an absolute element. In accordance with the gauge principles, we then find that a nontrivial nonmetricity $Q_{kij}\neq 0$ of spacetime emerges and the Riemann-Cartan geometry is extended to a metric-affine geometry with independent metric $g_{ij}$ and independent connection $\Gamma_{kj}{}^i$. 

In the simplest case, we could then use the trace of the nonmetricity, the {\it Weyl covector} $Q_k := {\frac 14}g^{ij}Q_{kij}$ as tool. This seems to be the minimal way to violate (homogeneous) Lorentz invariance. However, such a Weyl-Cartan spacetime, because of its dilation invariance, is expected to be valuable only in the context of massless fields. For massive fields we have to take recourse to the complete nonmetricity, including its trace-free (or deviatoric) part $\slash\hspace{-8pt} Q_{kij}:=Q_{kij}-Q_kg_{ij}$. In 4 dimensions, the 40 component nonmetricity tensor $Q_{kij}$ can be decomposed into four irreducible under the $GL(4,R)$ parts, according to $40=16\,\oplus 16\,\oplus 4\,\oplus 4$, see \cite[App.~B.1]{Hehl:1994ue}.

The metric-affine geometric formalism provides a natural and consistent description of the possible Lorentz symmetry violation. A look at the literature reveals, however, that the latter is mostly described in various non-geometric formulations with the help of non-dynamical Lorentz-violating (LV) tensor fields that may arise from nontrivial vacuum expectation values of appropriate quantum operators, see Mattingly \cite{Mattingly}, for instance, and Kiefer \cite[p.~330]{Kiefer}. Such LV tensors may be constructed in terms of a timelike vector field which is introduced as a new element of spacetime structure. Taking into account that gravity is universally interacting with matter, the non-dynamical LV fields are replaced with dynamical tensor fields in Einstein-Aether theories, for specific models see Eling et al \cite{Eling}, Balakin et al \cite{Balakin:2022zng}, and Jacobson \cite{Jacobson:2007veq}. It should be stressed, however, that in these models the spacetime geometry is kept Riemannian, but still a violation of Lorentz invariance is claimed to arise, see the discussions of the Standard Model Extension (SME) overviewed by Bailey \cite{Bailey:2023pfd}, Liberati \cite{Liberati:2013}, Kostelecky \cite{Kostelecky:2022idz}, and Heros et al. \cite{Perez}.

The naturalness of such LV mechanisms is an open issue \cite{Liberati:2013}, and the origin of an ad hoc Lorentz covariant vector field inducing the Lorentz symmetry violation is unclear. Quite paradoxically, such ad hoc non-geometric structures are still widely used to describe the possible violation of Lorentz invariance, instead of turning to intrinsic geometrical covector fields such as the Weyl field and, more generally, the nonmetricity field in the framework of the metric-affine gravity theory\footnote{See, though, the discussion of constraints on the post-Riemannian structures in the LV framework \cite{Foster:2016uui,Zhu:2023kjx}.}.

\section{Hyperfluid controlled by the energy-momentum and the hypermomentum laws}

In order to probe the spacetime structure in the framework of the metric-affine approach with independent metric and connection, one needs the matter with microstructure \cite{Hehl:1994ue,Neeman:1996}. 

Historically, the emancipation of the connection structure from its dominance by the metric structure proceeded in several steps: First Voigt (1887) \cite{Voigt}, in continuum mechanics, taking certain crystal lattices as a lead, introduced so-called (spin) moment stresses  $\tau_{ij}{}^k = -\,\tau_{ji}{}^k$ as a new concept besides the now asymmetric (force) stresses $\Sigma_{ij}\neq\Sigma_{ji}$. The brothers Cosserat (1909) \cite{Cosserat} developed a corresponding classical field theory, see also Truesdell and Toupin \cite{Truesdell}. As a result, for the first time the equilibrium conditions for moments, $\partial_k\tau_{ij}{}^k + \Sigma_{[ij]} = 0$, became independent from the corresponding conditions for forces, $\partial_k\Sigma_i{}^k = 0$. Or, 4-dimensionally speaking, angular momentum conservation became independent from energy-momentum  conservation. 

Further development of the continuous mechanics of media with microstructure \cite{Truesdell,capriz} resulted in spin fluid models \cite{WR,OK}, the dynamics of which is satisfactorily described in the framework of the Cosserat approach \cite{Cosserat,Halbwachs,Halbwachs2}. The elements of such media are characterized by a {\it rigid material frame}, representing the degrees of freedom of an intrinsic rotation, or spin, of matter elements, thereby giving rise to a spin density ${\mathcal S}_{ij} = -\,{\mathcal S}_{ji}$ tensor of the fluid. In contrast to the structureless ideal fluid of GR, such a medium with microstructure is described by {\it two} Noether currents: the canonical energy-momentum and the spin. For the {\it pressure-less} case, they read
\begin{equation}\label{SigW}
\Sigma_i{}^k = u^k{\mathcal P}_i,\qquad \tau_{ij}{}^k = u^k{\mathcal S}_{ij}.
\end{equation}
The 4-momentum ${\mathcal P}_k \neq {\frac {\rho}{c^2}}\,u_k$ is no longer collinear the 4-velocity, and it carries an additional spin-dependent contribution.
    
The hyperfluid model was developed \cite{Obukhov:1993} as a natural extension of the concept of a spin fluid to the case of when matter elements carry a {\it deformable material frame}, thus adding to the spin the intrinsic dilation and shear degrees of freedom to form the {\it hypermomentum density} ${\mathcal J}^i{}_j$ of a continuum. The resulting medium with microstructure is then described by the canonical energy-momentum and the hypermomentum (\ref{D})
\begin{equation}\label{SigJ}
\Sigma_i{}^k = u^k{\mathcal P}_i,\qquad \Delta^i{}_j{}^k = u^k{\mathcal J}^i{}_j,
\end{equation}
which generalize (again for the {\it pressure-less} case) the pair of Noether currents (\ref{SigW}). The spin density is identified with the skew-symmetric part ${\mathcal S}_{ij} = {\mathcal J}_{[ij]}$, whereas the {\it dilation density} is the trace ${\mathcal J} = {\mathcal J}^i{}_i$, and the symmetric traceless part is the {\it shear density}. 

The consistent variational theory of hyperfluid \cite{Obukhov:2023} yields the explicit form of the canonical energy-momentum tensor (\ref{SigJ}) for the case of nontrivial {\it pressure} $p$:
\begin{align}\label{Sig}
\Sigma_k{}^i &= u^i{\mathcal P}_k - p\left(\delta_k^i - {\frac {u_ku^i}{c^2}}\right),\\
{\mathcal P}_k &= {\frac {\rho}{c^2}}\,u_k - {\frac 1{c^2}}(g_{kj}u^l
- \delta_k^lu_j)\dot{\mathcal J}^j{}_l\,,\label{Pk}
\end{align}
and the equation of motion of the hypermomentum reads
\begin{equation}\label{eomJ}
\dot{\mathcal J}^i{}_j - {\frac {1}{c^2}}\,u^iu_k\dot{\mathcal J}^k{}_j - {\frac {1}{c^2}}
\,u_ju^k\dot{\mathcal J}^i{}_k + {\frac {1}{c^4}}\,u^iu_ju^lu_k\dot{\mathcal J}^k{}_l = 0.
\end{equation}
Here the dot denotes the substantial derivative along the fluid's flow $\dot{\mathcal J}{}^i{}_j := {\stackrel {*}\nabla}_k(u^k{\mathcal J}{}^i{}_j)$, where the modified covariant derivative is defined as ${\stackrel * \nabla}{}_i := {\stackrel \Gamma \nabla}{}_i - T_{ki}{}^k - {\frac 12}Q_{ik}{}^k$. 

The standard Euler-Lagrange machinery yields the conservation laws of the hypermomentum and the energy-momentum, respectively:
\begin{align}\label{tskew}
{\stackrel * \nabla}{}_j \Delta^i{}_k{}^j &= \Sigma_k{}^i - \sigma_k{}^i,\\
{\stackrel * \nabla}{}_i\Sigma_k{}^i &= \Sigma_l{}^i T_{ki}{}^l - \Delta^m{}_n{}^l R_{klm}{}^n
- {\frac 12}\sigma^{ij}Q_{kij}.\label{cons2b}
\end{align}

\section{Conclusions and outlook}

By definition, the Lorentz group consists of transformations of coordinates and frames that does change the spacetime metric with the Minkowski signature $(+,-,-,-)$. From gauge-theoretic principles, we then find the geometry without nonmetricity. However, when the metric loses its status of an absolute element and the Lorentz invariance in broken, the geometric structure of spacetime acquires nontrivial nonmetricity. In other words, the Lorentz symmetry violation and the nonmetricity go hand-in-hand.

Following Einstein, the geometry of spacetime is not fixed by postulates or conventions for the sake of simplicity and convenience, but it should rather be determined from physical observations and experiment. It is sufficient to use structureless test matter to probe the Riemannian geometry. However, one does need to employ test {\it matter with microstructure} to explore post-Riemannian geometries. Practical advice to experimentalist is as follows: use matter with intrinsic spin to detect torsion, and use matter with intrinsic hypermomentum to detect nonmetricity.

These conclusions are convincingly supported by the analysis of equations of motion of test bodies in post-Riemannian geometries, \cite{Yasskin} and \cite{Puetzfeld:2007,Obukhov:2015,Iosifidis:2023eom}. Derived from the most general conservation laws of the Noether currents, the equations of motion may technically look slightly different in different multipolar approximation schemes, but the qualitative conclusion above remains untouched. 

Quantum spin dynamics can be effectively used in the search for the torsion manifestations \cite{Shapiro:2002,Ni:2010,Obukhov:2014}. Still, the study of the quantum hypermomentum dynamics is an open issue, and in the meantime the development of physically feasible classical systems such as the hyperfluid model is of interest. The hyperfluid model attracted considerable attention in the analysis of the dynamics of micromorphic hyperelastic continua \cite{capriz:2002,capriz:2005,capriz:2007}, whereas in the gravity theory it was mostly used in the cosmological context \cite{Smalley:1995,Babourova,Puetzfeld:2001,Iosifidis:2020,Iosifidis:2021,Iosifidis:2022}.

An appropriate account for irreversible thermodynamic aspects, more general than those in  \cite{IngoMueller,JouLebon}, could pave way to possible applications of the hypermomentum concept and the hyperfluid model ranging from the early cosmology to the heavy ion physics \cite{Singh:2023}. The hyperfluid could be used as a classical approximation for the study of the quark-gluon plasma dynamics, where the use of a spin fluid (see Beccatini \cite{Becattini:2022zvf} and Biswas et al.\  \cite{Biswas:2022bht}) appears to be too restrictive, in our  opinion, and an account on the Regge-trajectories-like hadronic excitations is needed.

\smallskip
{\bf ACKNOWLEDGMENTS}

\vspace{-0.75mm}
We are grateful to Ingo M\"uller (Berlin) for useful explanations to `Extended Thermodynamics.'

\end{document}